# Nonlinear stimulated Brillouin scattering in a single-mode optical fiber

K. Kieu,[*] D. Churin, E. M. Wright, R. A. Norwood, and N. Peyghambarian

[1]*College of Optical Sciences, University of Arizona, Tucson, Arizona 85721*
*Corresponding author: kkieu@optics.arizona.edu*

We predict and experimentally observe a nonlinear variant of stimulated Brillouin scattering in a single-mode fiber that arises from consideration of higher-order processes for phonon generation. The effect manifests itself at high laser excitation as the appearance of Stokes gain for a detuning equal to half of the conventional Brillouin frequency in the fiber, with no accompanying anti-Stokes absorption at the opposite detuning, and requires counter-propagating pump beams for phase-matching. We believe that this could be a new nonlinear optical effect that has not been observed before.
© 2013 Optical Society of America

Brillouin and Raman scattering have played pivotal roles in the development of our understanding of light-matter interactions. Brillouin scattering was predicted by L. Brillouin in 1922 [1] and experimentally observed by E. Gross in 1930 using a lamp as the light source [2]. Spontaneous Brillouin scattering is normally viewed as a linear optical effect, as is spontaneous Raman scattering, since the underlying physical mechanism involves inelastic light scattering of pump light from thermally excited acoustic waves, or phonons, yielding weak conversion to a Stokes frequency-shifted wave. With the advent of lasers, stimulated Brillouin scattering (SBS) was observed for which the conversion efficiency can be 100% [3, 4]. In current fiber optics communication technology steps are often taken to avoid the detrimental effects of SBS, but interesting applications exist such as inertial rotation measurement [5], thermal and strain sensing [6], ultra-narrow linewidth light sources [7, 8], low noise microwave generation, and slow light [9, 10].

Stimulated Brillouin scattering also shares common features with the distinctly linear effect of acousto-optics, both involving scattering of incident pump light from an acoustic wave and generation of a Stokes-shifted wave. Indeed the induced polarization $\Delta P(\vec{r},t) = \epsilon_0 \Delta \chi(\vec{r},t) E(\vec{r},t)$ responsible for both is proportional to the product of the electric field times the acoustic-wave amplitude, the change in susceptibility $\Delta \chi(\vec{r},t)$ being proportional to the acoustic wave amplitude. The distinction between the two is that for acousto-optics the acoustic wave is externally defined, whereas in SBS the acoustic field is internally generated, via electrostriction, by the intensity interference between the incident optical field and the generated Stokes field. From this perspective SBS may be viewed as a linear coupled system in which the optical fields are modified by the acoustic field, and the acoustic field is in turn driven by the optical fields. Then owing to the positive feedback that can occur at high pumping, SBS behavior resembles that in a laser: at low pump power the generated Stokes power is small (spontaneous Brillouin scattering), whereas at and beyond a threshold pump power the output Stokes power increases, initially exponentially but then linearly as depletion of the pump occurs. As is well known, these features of SBS may be derived from the theory in which the acoustic wave is adiabatically eliminated and solved for in terms of the pump and Stokes fields, yielding a phenomenological description as a third-order nonlinear optical process. This nonlinear optics approach is the one usually found in textbooks and the current literature [11, 12].

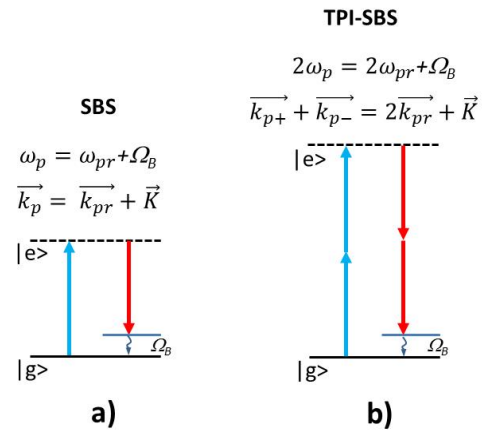

Figure 1: Energy level diagrams showing the transitions underlying phonon generation in standard SBS and the proposed NL-SBS. The corresponding equations governing conservation of energy and momentum are included above each diagram. In both cases the ground state is denoted |g> and the off-resonant excited states |e> are represented by a virtual level.

The aim of this Letter is to introduce and explore experimentally a nonlinear variant of SBS for which the induced polarization is nonlinear in the optical fields even before adiabatic elimination of the acoustic wave, putting this process on par with other direct nonlinear optical processes such as harmonic generation. This nonlinear SBS (NL-SBS) arises from consideration of higher-order processes for phonon generation, and it leads to distinct features in comparison to standard SBS. In particular, it leads to SBS gain/loss for a pump-probe detuning of half the usual Brillouin frequency, with no associated SBS gain/loss for the opposite detuning, and it requires counter-propagating pump beams to be phase-matched. All of these features, which are very different from standard SBS, are demonstrated experimentally using a single-mode optical fiber subject to high power and continuous-wave pump and probe beams. To the best of our knowledge this is the first report of NL-SBS.

To proceed we first discuss the basic ideas underlying standard SBS as a basis for introducing NL-SBS. As illustrated in Fig. 1.a, the quantum process underlying standard SBS involves annihilation of a pump photon followed by creation of both a Stokes photon and an acoustic phonon of frequency $\Omega_B$. Here the pump is labeled by a subscript 'p' and the probe by a subscript 'pr'. For SBS gain the probe frequency equals the Stokes-shifted frequency. Energy conservation for this process demands that $\omega_{pr} = \omega_p - \Omega_B$. The corresponding phase-matching condition, which follows from conservation of momentum, dictates that the generated Stokes photon must travel in the reverse direction along the fiber axis compared to the pump photon. The equation also tells us that the modulus of the wavevector of the acoustic phonon $K$ should be about twice the modulus of the wavevector of the pump photon, $K \approx 2k_p$, since $\omega_p \gg \Omega_B$.

In contrast to standard SBS the simplest higher-order quantum process underlying NL-SBS is shown in Fig. 1.b and involves annihilation of two pump photons followed by creation of two probe photons and an acoustic phonon of frequency $\Omega_B$ [13]. We term this specific process two-photon induced (TPI) SBS since it involves initial absorption of two pump photons. Energy and momentum conservation for this process leads to three distinct features of TPI-SBS that may be probed experimentally. First, energy conservation for this process demands that $(\omega_p - \omega_{pr}) = \Omega_B / 2$, that is, the pump-probe detuning is half the Brilloiun frequency. Second, since the phonon dispersion relation demands $K \approx 2k_p$, phase-matching of TPI-SBS requires counter-propagating pump beams, labeled $p \pm$, so that $\vec{k}_{p+} + \vec{k}_{p-} = 0$. Third, the anti-Stokes resonance that may be expected for $(\omega_p - \omega_{pr}) = -\Omega_B / 2$ is not assured based on symmetry alone. In particular, for standard SBS the anti-Stokes resonance follows from interchanging the roles of the pump and probe. Since the pump and probe are counter-propagating, and no direction is privileged over the other, the Stokes and anti-Stokes resonances should both appear together. This symmetry argument fails for TPI-SBS since there are counter-propagating pump fields but a unidirectional probe, thereby breaking the forward-backward symmetry, so it is possible to have one resonance without the other. If counter-propagating pumps and probes are employed both the Stokes and anti-Stokes resonances would be expected.

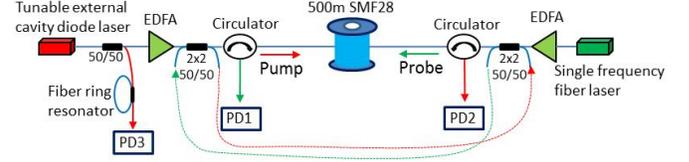

Figure 2: Schematic diagram of the experimental setup. PD1, PD2 and PD3 are photodiodes.

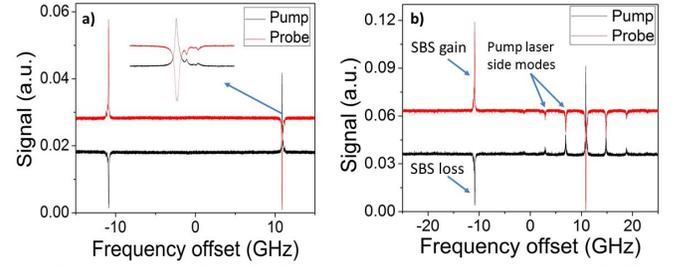

Figure 3: Standard Brillouin spectroscopy on 500m of SMF28 fiber. a) Low power measurement (Pump ~ 40 mW, Probe ~ 60 mW); b) Tunable laser (Pump) modes appear at elevated driving field strength (Pump ~ 70 mW, Probe ~ 125 mW). The measured Brillouin shift is 10.86 GHz. The frequency offset is the probe frequency minus the pump frequency.

Based on the above considerations we constructed the experimental setup shown in Fig. 2, which depicts a scheme to perform Brillouin spectroscopy with a unidirectional probe and counter-propagating pumps. This scheme is advantageous compared to the case where only the pump field is applied since we resonantly drive the acoustic phonons with two strong fields so the signal to noise ratio is improved. The probe laser is a single frequency fiber laser (Koheras) with <1 kHz linewidth. The pump laser is a tunable external cavity single frequency diode laser (Agilent 81680A). The spectral linewidth of the pump laser following the manufacturer's specification is < 100 kHz. We use home-built double-cladding pumped Er/Yb amplifiers to boost the output from the pump and probe lasers to > 1 W. A portion of the pump laser is split off and sent through a fiber ring resonator for frequency calibration purpose (the wavelength tuning speed of the Agilent laser is not uniform). The fiber ring resonator is made from a 98/2 2x2 coupler and has a free spectral range of ~95.3 MHz. In our experiments, we used a spool of SMF28 fiber with 500 m in length. The optimal length of the SMF28 fiber is chosen empirically so that the threshold of TPI-SBS is lower than that for spontaneous SBS, since once the threshold for spontaneous SBS is reached the noise on the photodiodes (1 and 2) is significantly increased due to the stochastic nature of spontaneous SBS. The two thresholds appeared to be at the same level for 1 km of SMF28 fiber. The two 2x2 50/50 fiber couplers on both side of the 500 m SMF28 fiber are used to launch the pump or probe in opposite directions. Two polarization controllers (not shown in Fig. 2) are used to optimize the polarization states of the pump and probe fields. PD1 is used to monitor the transmission of the probe signal; PD2 is used to monitor the transmission of the pump signal; and PD3 is used to record the longitudinal modes of the fiber ring resonator for frequency calibration purpose.

In the first experiment, we perform standard stimulated Brillouin spectroscopy where the pump propagates in one direction and the probe propagates in the opposite direction. By sweeping the Pump laser frequency across the Probe laser frequency (which is kept fixed) and monitoring the signals on PD1-PD3, the effect of SBS was observed at locations where the difference between the pump and probe laser frequencies matched the Brillouin resonance of the silica glass fiber (Fig. 3(a)). Amplification was observed when the probe laser frequency was on the Stokes side and stimulated Brillouin absorption was observed on the anti-Stokes side, as expected. It turned out that our Agilent tunable laser has some small sidebands from other longitudinal modes (4GHz spacing). These modes carried a small amount of energy but could be amplified to significant levels in the present of the probe beam when the probe beam's frequency is higher than the pump's frequency (Fig. 3, right). Note that the role of pump and probe signals is essentially interchangeable in this case since the response is symmetric with regard to the frequency offset (which in our notation is defined as the frequency difference between the probe and the pump signals).

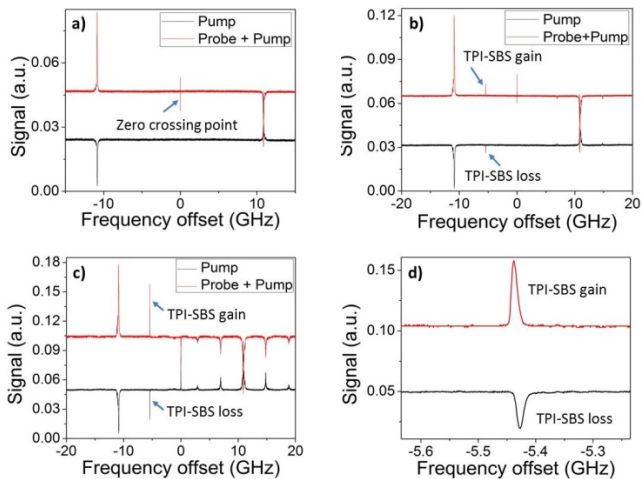

Figure 4: Observation of TPI-SBS. Two pump fields are launched in opposite directions. The TPI-SBS gain and loss are observable above some threshold level (around 50 mW for all fields). a) probe ~ 40 mW, pump ~ 40 mW (in both directions); b) probe ~ 60 mW, pump ~ 60 mW (in both directions); c) probe ~ 100 mW, pump ~ 100 mW (in both directions); d) zoom-in to the NL-SBS gain and absorption lines from c).

To observe TPI-SBS we need to have two pump fields propagating in opposite directions as discussed above (Fig. 1.b). To create the required condition, a portion of the pump signal (split off from the 2x2 coupler) is launched in the opposite direction (along with the probe signal) using the available port of the other 2x2 coupler on the probe laser side (red dotted line). PD1 now detects both the probe and the pump signal going from right to left. PD2 still detects only the pump signal going from left to right. Now, by tuning the pump laser wavelength (40 nm/s) we can observe the TPI-SBS gain or loss in the same way as was done for standard SBS. However we find that the TPI-SBS resonance now occurs at exactly half the Brillouin frequency as predicted above (at 5.43 GHz, as shown in Fig. 4). Furthermore, we find that TPI-SBS occurs only when the pump frequency is higher than the probe frequency (negative frequency offset), again in agreement with theoretical expectations. To further confirm our predictions we considered the case with counter-propagating probes and a unidirectional pump (using the green dotted fiber line in Fig. 2). The measurement result is shown in Fig. 5. Here, we also observed TPI-SBS at half the Brillouin frequency but the location of the induced gain and loss resonant lines are now on the positive frequency offset as expected.

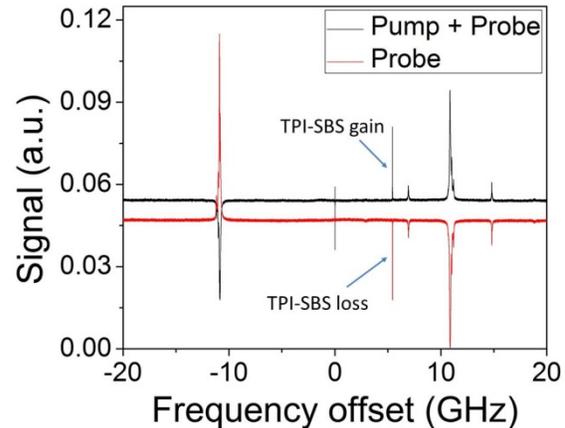

Figure 5: Two probe fields are launched in opposite directions. The TPI-SBS induced gain and loss are observed as well but now on the positive frequency offset side as expected. Pump ~ 30 mW; Probe ~ 90 mW (both directions).

So far our discussion of TPI-SBS has involved general features arising from energy and momentum conservation. We do not yet have a fully quantitative theory but below we offer some ideas regarding the underlying physics. We start from the acoustic wave equation for small density variations $\tilde{\rho}(\vec{r},t)$ in the fiber [11,12]

$$\left(\frac{\partial^2}{\partial t^2} - \Gamma'\nabla^2\frac{\partial}{\partial t} - v^2\nabla^2\right)\tilde{\rho} = -\frac{1}{2}\nabla^2(\varepsilon_0\chi|E|^2), \quad (1)$$

where $\rho_0 \gg |\tilde{\rho}|$ is the mean density, $v$ is the speed of sound, $\Gamma'$ is the damping parameter, $\chi$ is the medium susceptibility, and $E$ is the positive frequency component of the electric field with factor $\exp(-i\omega_{pr}t)$ removed, $\omega_{pr}$ being the probe frequency. The right-hand side of Eq. (1) describes coupling between the density variations and the laser fields via electrostriction [12]. First we review the physics underlying standard SBS. Consider the general case with input field $E = (E_{p+} + E_{p-} + E_{pr})$, with two pump fields of amplitudes $E_{p\pm} = A_{p\pm}e^{\pm ikz - i\Omega t}$ counter-propagating along the z-axis along with a backward propagating probe field $E_{pr} = A_{pr}e^{-ikz}$, $\Omega = (\omega_p - \omega_{pr})$ being the detuning of the pump with respect to the probe. Then for a forward propagating pump beam ($E_{p-} = 0$) the term $\chi|E|^2$ in Eq. (1) contains a driving term for the acoustic wave $\tilde{\rho} \propto A_{p+}^*A_{pr}e^{i\Omega t - 2ikz}$ (along with the complex conjugate term) that can resonantly excite a forward propagating acoustic wave if the detuning is equal to the Brillouin frequency $\Omega = \Omega_B = 2vn\omega_p/c$ [12]. Subsequent scattering of forward propagating pump photons from the acoustic wave provides an induced polarization $\Delta P \propto \tilde{\rho}E_{p+}$ at the probe frequency that can produce gain for the Stokes wave.

We propose that TPI-SBS involves physics beyond the standard theory of electrostriction in Eq. (1). In particular, above we tacitly assumed that the susceptibility is a constant whereas in general it can contain a nonlinear contribution $\chi = \chi_0 + \chi^{(3)}|E|^2$, with $\chi^{(3)}$ the third-order electronic nonlinear susceptibility. Thus, the right hand side of Eq. (1) also involves the term $\nabla^2(\varepsilon_0\chi^{(3)}|E|^4)$. In particular, including both counter-propagating pump beams, the new term allows a driving term for the acoustic wave $\tilde{\rho} \propto A_+^*A_-^*A_{pr}^2 e^{2i\Omega t - 2ikz}$ (along with the complex conjugate term) that, analogous to the standard case, can resonantly excite a forward propagating acoustic wave but with two caveats: First this model leads naturally to the condition $\Omega = \Omega_B/2$ observed in the experiment. Second, excitation of the acoustic wave requires both pump beams to be present, again in agreement with the experiment to the extent that TPI-SBS vanishes with only one pump beam present. Subsequent scattering of one of each of the pump and probe photons from the acoustic wave provides an induced polarization $\Delta P \propto \tilde{\rho}E_{p+}E_{p-}E_{pr}^*$ at the probe frequency that can produce gain for the Stokes wave: This form of the induced polarization follows from setting up the quantum Hamiltonian for TPI-SBS portrayed in Fig. 1.b and evaluating the Heisenberg equation of motion for the Stokes field [14]. This analysis suggests that TPI-SBS may be viewed not only as due to inclusion of higher-order processes, as illustrated in Fig. 1, but also as a correction to the usual theory of electrostriction underlying SBS.

In conclusion, we have introduced and explored the new idea of nonlinear stimulated Brillouin scattering that arises from considering higher-order processes for phonon generation. In particular, we predicted and experimentally demonstrated TPI-SBS occurring at half of the conventional Brillouin frequency, and we argued that this points to a modification of the standard theory of electrostriction. To the best of our knowledge this is the first ever experimental evidence of such corrections to the process of SBS in dielectric media. By generalizing our view to other similar nonlinear processes we conjecture that one may observe SBS at harmonics of the conventional Brillouin shift with the right phase matching condition. Finally, we expect that our result may be applied to Raman scattering in a similar fashion.

example different time ordering. We concentrate on the specific case in Fig. 1.b as it is the most transparent.
14. A. Yariv, IEEE J. Quant. Electron **1**, 28-36 (1965).